\documentclass{article}
\usepackage{amsfonts}
\newcommand{\be}{\begin{equation}}
\newcommand{\ee}{\end{equation}}
\newcommand{\bea}{\begin{eqnarray}}
\newcommand{\eea}{\end{eqnarray}}
\newcommand{\beaa}{\begin{eqnarray*}}
\newcommand{\eeaa}{\end{eqnarray*}}

\newcommand{\BB}{{{\rm I} \kern -2pt \rlap {\rm B} \kern +8pt}}

\advance\textheight by \topskip
\makeatletter

\@addtoreset{equation}{section}
\def\section{\@startsection {section}{1}{\z@}{-3.5ex plus -1ex minus
 -.2ex}{2.3ex plus .2ex}{\large\bf\centering}}
\def\subsection{\@startsection{subsection}{2}{\z@}{-3.25ex plus -1ex minus -.2ex}{1.5ex plus .2ex}{\bf}}
\def\subsubsection{\@startsection{subsubsection}{3}{\z@}{-3.25ex plus -1ex minus -.2ex}{1.5ex plus .2ex}{\sl}}
\makeatother

\begin{document}

\baselineskip 18pt \parindent 12pt \parskip 10pt

\begin{titlepage}

\begin{center}
{\Large {\bf On algebraic structures in supersymmetric principal chiral model }}\\\vspace{1.5in} {\large Bushra Haider\footnote{%
bushrahaider@hotmail.com}  and M. Hassan \footnote{%
mhassan@physics.pu.edu.pk} }\vspace{0.15in}

{\small{\it Department of Physics,\\ University of the Punjab,\\
Quaid-e-Azam Campus,\\Lahore-54590, Pakistan.}}
\end{center}

\vspace{1cm}
\begin{abstract}
Using the Poisson current algebra of the supersymmetric principal
chiral model, we develop the algebraic canonical structure of the
model by evaluating the fundamental Poisson bracket of the Lax
matrices that fits into the $r-s$ matrix formalism of
nonultralocal integrable models. The fundamental Poisson bracket
has been used to compute the Poisson bracket algebra of the
monodromy matrix that gives the conserved quantities in
involution.
\end{abstract}
\vspace{1cm} PACS: 11.30.Pb, 02.30.Ik
\end{titlepage}

\section{Introduction}
In some recent investigations \cite{1}, \cite{2}, the classical
integrability of supersymmetric principal chiral model (SPCM) has
been studied. It has been shown in $\cite{2}$, that for the SPCM
there exist two families of local conserved quantities in
involution, each with finitely many members whose spins are
exactly the exponents of the underlying Lie algebra of the model,
with no repetition modulo the Coxeter number. Along with the
existence of local conserved quantities in SPCM, there exist
non-local conserved quantities as well
\cite{1}-\cite{Curtright}$.$ Furthermore, it has been shown in
$\cite{1}$ that the SPCM also admits a one-parameter family of
transformation on superfields leading to a superfield Lax
formalism and a zero-curvature representation. The superfield Lax
formalism is shown to be related to the super Backlund
transformation and the super Riccati equations of the SPCM.

In integrable field theories some integrable canonical structures
are associated with the Lax pair. The Lax pair in general is a
pair of matrices, which are functions of fields and a spectral
parameter. The matrices obey a Poisson bracket algebra which in
many cases is ultralocal i.e. the algebra does not contain the
derivatives of the delta function. Such models are referred to as
ultralocal models. The ultralocality leads to a Poisson bracket
algebra of monodromy matrix and the Jacobi identity gives the
classical Yang-Baxter equation for an
$r$-matrix\cite{4}-\cite{19}$.$ In fact, the Yang-Baxter equation
leads to the existence of commuting conserved quantities, ensuring
the integrability of the model\cite{4}-\cite{19}$.$

The $r$-matrix method has also been employed to non-ultralocal
models i.e. the models for which the algebra of Lax matrices
contains derivatives of the delta function $\cite{4}-\cite{19}$.
Some examples of such integrable models are the principal chiral
model (PCM), the complex sine-Gordon theory (CSG), the
Wess-Zumino-Witten model (WZW), $O\left( N\right) $ sigma model,
etc. The $r-s,$ matrix approach has been adopted to study such
non-ultralocal models in which the $r$-matrices are no longer
anti-symmetric and may also depend on dynamical variables giving
an extended dynamical Yang-Baxter equation.$\cite{9}.$

The purpose of this paper is to study the integrability of the
supersymmetric principal chiral model (SPCM) as a non-ultralocal
model, using the $r-s$ matrix approach of the Poisson bracket
algebra of monodromy matrix. We demonstrate that the SPCM which is
known to be integrable provide an explicit realization of the
$r-s$ matrix formalism developed for bosonic integrable models by
Maillet \cite{15}, \cite{16}. Starting with a Lax formalism, we
develop a canonical $r-s$ matrix approach for the SPCM and obtain
the Poisson bracket algebra of the monodromy matrix in terms of
the $r-s$ matrices for which the consistency condition implies an
extended non-dynamical Yang-Baxter equation.

\section{The SPCM and its Poisson bracket algebra}
Following \cite{1}, \cite{2}, we define the supersymmetric
principal chiral model as follows. Let us consider a superfield
$G\left( x,\theta \right) $ with values in a Lie group
$\mathcal{G}.$ This superfield $G\left( x,\theta \right) $ is a
function of the space coordinates $x^{\pm }$ and
anti-commuting coordinates $\theta ^{\pm }$ $\footnote{%
The orthonormal coordinates $x^{0}=t$ and $x^{1}=x$ in two
dimensions are
related to the light-cone coordinates and derivatives as $x^{\pm }=\frac{1}{2%
}\left( t\pm x\right) $, and $\partial _{\pm }=\partial _{t}\pm
\partial
_{x}.$}.$ The superspace Lagrangian of the SPCM is then given as%
\begin{equation}
\mathcal{L}=\frac{1}{2}\mbox{Tr}\left( D_{+}G^{-1}D_{-}G\right) ,
\label{lagrangian}
\end{equation}%
where
\begin{equation}
D_{\pm }=\frac{\partial }{\partial \theta ^{\pm }}-i\theta ^{\pm
}\partial _{\pm }  \label{covariant derivative}
\end{equation}%
are the superspace covariant derivatives and
\begin{equation}
G\left( x,\theta \right) G^{-1}\left( x,\theta \right)
=1=G^{-1}\left( x,\theta \right) G\left( x,\theta \right) .
\label{constraints}
\end{equation}%
The superspace Lagrangian $\mathcal{L}$ is invariant under the
following
transformation%
\begin{equation}
\mathcal{G}_{L}\times \mathcal{G}_{R}:\mbox{ \ \ \ \ \ \ \ \ \ \ \
}G\left( x^{\pm },\theta ^{\pm }\right) =\mathcal{U}\mbox{\
}G\left( x^{\pm },\theta ^{\pm }\right) \mathcal{V}^{-1},
\label{transformation}
\end{equation}%
where $\mathcal{U}$ and $\mathcal{V}$ are $\mathcal{G}_{L}$ and $\mathcal{G}%
_{R}$ valued matrix superfields respectively. The Noether
conserved superfield currents associated with the global
transformation are
\begin{equation}
J_{\pm }^{L}=iD_{\pm }GG^{-1},\mbox{ \ \ \ \ \ \ \ \ \ \ \ }J_{\pm
}^{R}=-iG^{-1}D_{\pm }G,  \label{global transformation}\
\end{equation}%
where $J_{\pm }^{R,L}$ are the Grassmann odd and are Lie algebra
valued, i.e., $J_{\pm }=J_{\pm }^{a}T^{a},$ where $\left\{
T^{a}\right\} $ is the
set of generators of the Lie algebra $\mathfrak{g}$ of the Lie group ${\cal G}$%
\footnote{%
Our Lie algebra conventions are as follows. The anti-hermitian generators $%
\left\{ T^{a},\,\,a=1,2,\ldots ,n=\dim \mathfrak{g}\right\} $ of
the Lie
algebra $\mathfrak{g}$ obey $\left[ T^{a},T^{b}\right] =f^{abc}T^{c}$ and Tr$%
\left( T^{a}T^{b}\right) =\delta ^{ab}.$ For any $X\in \mathfrak{g}$, $%
X=X^{a}T^{a}.$}. The superfield equation of motion of the SPCM is
the superfield conservation equation
\begin{equation}
D_{+}J_{-}^{R,L}-D_{-}J_{+}^{R,L}=0,  \label{eq of motion}
\end{equation}%
and the superfield zero curvature condition is identically
satisfied by $J_{\pm }$ as
\begin{equation}
D_{-}J_{+}^{R,L}+D_{+}J_{-}^{R,L}+i\left\{
J_{+}^{R,L},J_{-}^{R,L}\right\} =0, \label{zc condition}
\end{equation}%
 We can expand the
superfield $G\left(
x^{\pm },\theta ^{\pm }\right) $ as%
\begin{equation}
G\left( x,\theta \right) =g\left( x\right) \left( 1+i\theta
^{+}\psi^{R} _{+}\left( x\right) +i\theta ^{-}\psi^{R} _{-}\left(
x\right) +i\theta ^{+}\theta ^{-}F^{R}\left( x\right) \right) ,
\label{superfield expansion}
\end{equation}%
An alternative component expansion is given as%
\begin{equation}
G\left( x,\theta \right) = \left(i\theta ^{+}\psi^{L} _{+}\left(
x\right) +i\theta ^{-}\psi^{L} _{-}\left( x\right) +i\theta
^{+}\theta ^{-}F^{L}\left( x\right) \right)g\left( x\right)  ,
\end{equation}%
 where $\psi _{\pm
}$ are the Majorana spinors such that%
\begin{equation}
\psi^{R} _{\pm} = g^{-1}\psi^{L} _{\pm}g ,
\end{equation}%
 and $F\left( x\right) $ is
the
auxiliary field, with an algebraic equation of motion. The Majorana spinors $%
\psi _{\pm }\left( x\right) $ take values in the Lie algebra
$\mathfrak{g}$ of $\mathcal{G}$. The action of the symmetry $\mathcal{G}_{L}\times \mathcal{%
G}_{R}$ on component fields and the superfield currents $J _{\pm
}^{R,L}$ is
\begin{eqnarray*}
g &\longmapsto &UgV^{-1}, \\
\psi _{\pm }^{R} &\longmapsto &V\psi _{\pm }^{R}V^{-1}, \\
\psi _{\pm }^{L} &\longmapsto &U\psi _{\pm }^{L}U^{-1}, \\
J _{\pm }^{R} &\longmapsto &VJ _{\pm }^{R}V^{-1}, \\
J _{\pm }^{L} &\longmapsto &UJ _{\pm }^{L}U^{-1},
\end{eqnarray*}%
where $U$ and $V$ are the leading bosonic components of the matrix
superfields $\mathcal{U}$ and $\mathcal{V}$ respectively, i.e.,
the fermions transform under $\mathcal{G}_{R}.$
From now on we consider the spinors and current corresponding to $%
\mathcal{G}_{R}$, i.e., $\psi_{\pm }^{R}$ and $J_{\pm }^{R}$
(which we write as $\psi_{\pm }$ and $J_{\pm }$ during further
discussion).

 After the
elimination of the auxiliary field from the expression of $G\left(
x,\theta \right)$ , the component
Lagrangian finally becomes\footnote{%
The two-dimensional Minkowski matrix is $\eta _{\mu \nu }=\left(
\begin{array}{cc}
1 & 0 \\
0 & -1%
\end{array}%
\right) $ and the $\gamma $-matrices $\gamma _{0}=\left(
\begin{array}{cc}
0 & i \\
-i & 0%
\end{array}%
\right) ,$ $\gamma _{1}=\left(
\begin{array}{cc}
0 & i \\
i & 0%
\end{array}%
\right) $ satisfy $\left\{ \gamma _{\mu },\gamma _{\nu }\right\}
=2\eta _{\mu \nu }.$ The Dirac spinor is $\psi =\left(
\begin{array}{c}
\psi _{+} \\
\psi _{-}%
\end{array}%
\right) ,$ where $\psi _{\pm }$ are chiral spinors and our
assumption is
that $\psi _{\pm }$ are real (Majorana). The Lorentz behaviour of $x^{\pm }$,%
$\partial _{\pm }$ and $\psi _{\pm }$ is $x^{\pm }\mapsto e^{\mp
\Lambda
}x^{\pm },$ $\partial _{\pm }\mapsto e^{\mp \Lambda }\partial _{\pm }$ and $%
\psi _{\pm }\mapsto e^{\mp \frac{1}{2}\Lambda }\psi _{\pm },$
where $\Lambda $ is the rapidity of the Lorentz boost. The rule of
raising and lowering
spinor indices is $\psi ^{\pm }=\pm \psi _{\mp }.$}%
\begin{eqnarray}
\mathcal{L} &=&\frac{1}{2}\mbox{Tr}(g^{-1}\partial _{+}gg^{-1}\partial _{-}g  \nonumber\\
&&+i\psi _{+}\left( \partial _{-}\psi _{+}+\frac{1}{2}\left[
g^{-1}\partial
_{-}g,\psi _{+}\right] \right)  \nonumber \\
&&+\frac{i}{2}\psi _{-}\left( \partial _{+}\psi
_{-}+\frac{1}{2}\left[
g^{-1}\partial _{+}g,\psi _{-}\right] \right)   \nonumber\\
&&+\frac{1}{2}\psi _{+}^{2}\psi _{-}^{2}).  \label{compt
lagrangian}
\end{eqnarray}%
Using Euler-Lagrange equations, we can directly find the component
equations of motion for the SPCM. From equation $\left(
\ref{global transformation}\right)
$ we write the component expansion of superfield current of the SPCM as%
\begin{eqnarray}
J_{\pm } &=&\psi _{\pm }+\theta ^{\pm }j_{\pm }-\frac{i}{2}\theta
^{\mp
}\left\{ \psi _{+},\psi _{-}\right\}  \nonumber \\
&&-i\theta ^{+}\theta ^{-}\left( \partial _{\pm }\psi _{\mp
}-\left[ j_{\pm },\psi _{\mp }\right] -\frac{i}{2}\left[ \psi
_{\pm }^{2},\psi _{\mp }\right] \right) ,
\end{eqnarray}%
where the components of the bosonic current are given by%
\begin{equation}
j_{\pm }=-\left( g^{-1}\partial _{\pm }g+i\psi _{\pm }^{2}\right)
. \label{bosonic current}
\end{equation}%
Again $j_{\pm }$ represent the right bosonic current $j_{\pm
}^{R}$. Substituting these into the superspace equations of
motion, collecting terms and writing $h_{\pm }=\psi _{\pm
}^{2}\Leftrightarrow h_{\pm }^{a}=\frac{1}{2}f^{abc}\psi _{\pm
}^{b}\psi _{\pm }^{c}$,
 we get the equations of motion for
fermionic and bosonic fields of the SPCM,\ \ \ \ \ \ \ \ \ \ \ \ \
\ \ \ \ \ \ \ \ \ \ \ \ \ \ \ \ \ \ \ \ \ \ \ \ \ \ \ \ \ \ \ \
\begin{equation}
\partial _{\pm }\psi _{\mp }-\frac{1}{2}\left[ j_{\pm },\psi _{\mp }\right] -%
\frac{i}{4}\left[ h_{\pm },\psi _{\mp }\right] =0\noindent \ ,
\label{s1}
\end{equation}

\begin{equation}
\partial _{-}j_{+}+\partial _{+}j_{-}=0\noindent ,\ \ \ \   \label{s2}
\end{equation}%
along with

\begin{equation}
\partial _{\mp }j_{\pm }=-\frac{1}{2}\left[ j_{\pm },j_{\mp }\right] +\frac{i%
}{4}\left[ j_{\mp },h_{\pm }\right] -\frac{i}{4}\left[ j_{\pm },h_{\mp }%
\right] +\frac{1}{4}\left[ h_{\pm },h_{\mp }\noindent \right] \ .
\label{s3}
\end{equation}%
We use the fermion equations of motion to get the following
equations\ \ \ \ \ \ \ \ \ \ \ \ \
\begin{equation}
\partial _{-}j_{+}-\partial _{+}j_{-}+\left[ j_{+},j_{-}\right] =i\partial
_{-}h_{+}-i\partial _{+}h_{-},\noindent \   \label{s4}
\end{equation}

\begin{equation}
\partial _{\mp }(ih_{\pm })=-\frac{1}{2}\left[ ih_{\pm },j_{\mp }+\frac{i}{2}%
h_{\mp }\right] ,  \label{s5}
\end{equation}%
and%
\begin{equation}
\partial _{-}h_{+}+\partial _{+}h_{-}=0.\noindent  \label{h1}
\end{equation}%
The equations (\ref{s2}) and (\ref{h1}) show the conservation of
bosonic currents $j_{\pm }$ and $h_{\pm }$ respectively.

\smallskip The Poisson brackets for the bosonic currents have already been
derived in \cite{2} and are given below

\begin{eqnarray}
\left\{ j_{0}^{a}\left( x\right) ,j_{0}^{b}\left( y\right)
\right\}
&=&f^{abc}j_{0}^{c}\left( x\right) \delta \left( x-y\right) ,  \nonumber \\
\left\{ j_{0}^{a}\left( x\right) ,j_{1}^{b}\left( y\right)
\right\} &=&f^{abc}j_{1}^{c}\left( x\right) \delta \left(
x-y\right) +\delta
^{ab}\delta ^{\prime }\left( x-y\right) ,  \nonumber \\
\left\{ j_{1}^{a}\left( x\right) ,j_{1}^{b}\left( y\right) \right\} &=&\frac{%
-i}{4}f^{abc}(h_{+}^{c}\left( x\right) +h_{-}^{c}\left( x\right)
)\delta \left( x-y\right) .
\end{eqnarray}%
In light-cone coordinates, the brackets are expressed as

\begin{eqnarray}
\left\{ j_{\pm }^{a}\left( x\right) ,j_{\pm }^{b}\left( y\right) \right\} &=&%
\frac{1}{2}f^{abc}(3j_{\pm }^{c}\left( x\right) -j_{\mp
}^{c}\left( x\right) -\frac{1}{2}ih_{+}^{c}\left( x\right)
\nonumber \\
&&-\frac{1}{2}ih_{-}^{c}\left( x\right) )\delta \left(
x-y\right)+2\delta
^{ab}\delta ^{\prime }\left( x-y\right) , \\
\left\{ j_{+}^{a}\left( x\right) ,j_{-}^{b}\left( y\right) \right\} &=&\frac{%
1}{2}f^{abc}(j_{+}^{c}\left( x\right) -j_{-}^{c}\left( x\right) +\frac{1}{2}%
ih_{+}^{c}\left( x\right)  \nonumber \\
&&+\frac{1}{2}ih_{-}^{c}\left( x\right) )\delta \left( x-y\right)
,
\end{eqnarray}%
The fermions obey

\begin{eqnarray}
\left\{ \psi _{\pm }^{a}\left( x\right) ,\psi _{\pm }^{b}\left(
y\right) \right\} &=&-i\delta^{ab} \delta \left( x-y\right) ,
\\ \left\{ \psi _{+}^{a}\left( x\right) ,\psi _{-}^{b}\left(
y\right) \right\} &=&0.
\end{eqnarray}%
It is also useful to note that

\begin{eqnarray}
\left\{ h_{\pm }^{a}\left( x\right) ,\psi _{\pm }^{b}\left(
y\right) \right\} &=&if^{abc}\psi _{\pm }^{c}\left( x\right)
\delta \left( x-y\right)
, \\
\left\{ h_{\pm }^{a}\left( x\right) ,h_{\pm }^{b}\left( y\right)
\right\} &=&if^{abc}h_{\pm }^{c}\left( x\right) \delta \left(
x-y\right) .
\end{eqnarray}%
We recall the definition of the standard Poisson structure
associated with an arbitrary connected Lie group $\mathcal{G}$ and
consider a non-degenerate matrix $d$ with entries \cite{8},
\cite{Humphreys}

\begin{equation}
d^{ab}=\left\langle T^{a},T^{b}\right\rangle .  \label{c27}
\end{equation}%
where $<,>$ represents the killing form and for a semi-simple Lie
algebra we have $<T^{a},T^{b}>=\delta^{ab}$. Since
$Tr(T^{a}T^{b})=\delta^{ab}$, therefore if $\mathfrak{g}$ is a
semi-simple and represented as a matrix algebra, we may assume \
$d^{ab}=$Tr$(T^{a}T^{b}).$ Let $d_{ab}$ denote the entries of the
inverse matrix $d^{-1},$ an element $c $ of $\mathfrak{g}\otimes
\mathfrak{g}$ and elements $A^{a}$ of $\mathfrak{g}$ be defined by

\begin{eqnarray}
c &=&d_{ab}T^{a}\otimes T^{b},  \label{c28} \\[0.2in]
A_{a} &=&d_{ab}T^{b}.  \label{c29}
\end{eqnarray}%
Then we have the relations

\begin{equation}
\left[ c ,A_{c}\otimes I\right] =-\left[ c ,I\otimes A_{c}\right]
=f^{abc}A_{a}\otimes A_{b},  \label{c30}
\end{equation}%
where the symbols $A\otimes I$ \ and $I\otimes A$ \ denote the
natural embedding of $A$ into $\mathfrak{g}\otimes \mathfrak{g}$.

Using the usual tensor product notation, the Poisson brackets can
be
expressed in the following way:%
\begin{eqnarray}
\left\{ j_{0}\left( x\right)\stackrel{\otimes}{\mbox,} j_{0}\left(
y\right) \right\} &=&\left[ c ,j_{0}\left( x\right) \otimes
\mathbf{1}\right]
\delta \left( x-y\right) ,  \label{CR1} \\
\left\{ j_{0}\left( x\right) \stackrel{\otimes}{\mbox,}j_{1}\left(
y\right) \right\} &=&\left[ c ,j_{1}\left( x\right) \otimes
\mathbf{1}\right]
\delta \left( x-y\right)  \nonumber \\
&&+c \delta ^{\prime }\left( x-y\right) ,  \label{CR2} \\
\left\{ j_{1}\left( x\right) \stackrel{\otimes}{\mbox,}j_{1}\left(
y\right) \right\} &=&-\frac{i}{4}\left[ c ,\left( h_{+}\left(
x\right) +h_{-}\left( x\right) \right) \otimes \mathbf{1}\right]
\delta \left( x-y\right) ,
\label{CR3} \\
\left\{ j_{1}\left( x\right) \stackrel{\otimes}{\mbox,}h_{\pm
}\left( y\right) \right\} &=&\pm \frac{1}{2}\left[ c ,h_{\pm
}\left( x\right) \otimes
\mathbf{1}\right] \delta \left( x-y\right) ,  \label{CR4} \\
\left\{ j_{0}\left( x\right)\stackrel{\otimes}{\mbox,}h_{\pm
}\left( y\right) \right\} &=&\left[ c ,h_{\pm }\left( x\right)
\otimes \mathbf{1}\right]
\delta \left( x-y\right) ,  \label{CR5} \\
\left\{ h_{\pm }\left( x\right) \stackrel{\otimes}{\mbox,}h_{\pm
}\left(
y\right) \right\} &=&i\left[ c ,h_{\pm }\left( x\right) \otimes \mathbf{1}%
\right] \delta \left( x-y\right) ,  \label{CR6} \\
\left\{ h_{\pm }\left( x\right) \stackrel{\otimes}{\mbox,}h_{\mp
}\left( y\right) \right\} &=&0.  \label{CR7}
\end{eqnarray}%
In light-cone coordinates the above brackets can be expressed as%
\begin{eqnarray}
\left\{ j_{\pm }\left( x\right) \stackrel{\otimes}{\mbox,} j_{\pm
}\left( y\right) \right\} &=&\frac{1}{2}[c ,(3j_{\pm }\left(
x\right) -j_{\mp
}\left( x\right)  \nonumber \\
&&-\frac{1}{2}ih_{+}\left( x\right) -\frac{1}{2}ih_{-}\left(
x\right)
)\otimes \mathbf{1]}\delta \left( x-y\right)+2c \delta ^{\prime }\left( x-y\right) ,  \label{lcj11} \\
\left\{ j_{+}\left( x\right) \stackrel{\otimes}{\mbox,}
j_{-}\left( y\right) \right\} &=&\frac{1}{2}[c ,(j_{+}\left(
x\right) -j_{-}\left( x\right)
\nonumber \\
&&+\frac{1}{2}ih_{+}\left( x\right) +\frac{1}{2}ih_{-}\left(
x\right)
)\otimes \mathbf{1]}\delta \left( x-y\right) ,  \label{lcj111} \\
\left\{ j_{+}\left( x\right) \stackrel{\otimes}{\mbox,} h_{\pm
}\left( y\right)
\right\} &=&\frac{3}{2}\left[ c ,(h_{\pm }\left( x\right) \otimes \mathbf{1%
}\right] \delta \left( x-y\right) ,  \label{lcj2} \\
\left\{ j_{-}\left( x\right) \stackrel{\otimes}{\mbox,} h_{\pm
}\left( y\right)
\right\} &=&\frac{1}{2}\left[ c ,(h_{\pm }\left( x\right) \otimes \mathbf{1%
}\right] \delta \left( x-y\right) .  \label{lcj3}
\end{eqnarray}%
The SPCM is superconformally invariant classically, with the super
energy-momentum tensor obeying
\begin{eqnarray}
D_{-}\mbox{Tr}\left( J_{+}J_{+\mbox{ }+}\right) &=&0,  \label{ep1} \\
D_{+}\mbox{Tr}\left( J_{-}J_{-\mbox{ }-}\right) &=&0,  \label{ep2}
\end{eqnarray}%
where $J_{_{\pm \mbox{ }\pm }}$ is defined as%
\begin{equation}
J_{\pm \mbox{ }\pm }=D_{_{\pm }}J_{_{\pm }}+iJ_{_{\pm }}^{2}.
\label{ep3}
\end{equation}%
The component content of the superspace conservation equations
(\ref{ep1}) and (\ref{ep2}) correspond to the conservation of
supersymmetry current and the energy-momentum tensor $T_{_{\pm
\mbox{ }\pm }}.$ The conservation
equations for $T_{_{\pm \mbox{ }\pm }}$ are%
\begin{eqnarray*}
\partial _{+}T_{-\mbox{ }-} &=&0, \\
\partial _{-}T_{+\mbox{ }+} &=&0,
\end{eqnarray*}%
where the components $T_{_{\pm \mbox{ }\pm }}$ are given by%
\begin{equation}
T_{_{\pm \mbox{ }\pm }}=\mbox{Tr}\left( i\psi _{\pm }\partial
_{\pm }\psi _{\pm }+j_{\pm }^{2}+ij_{\pm }\psi _{\pm }^{2}\right)
. \label{ep4}
\end{equation}%
The Poisson brackets of the energy-momentum tensor components
$T_{_{\pm
\mbox{ }\pm }}$ are given as%
\begin{eqnarray}
\left\{ T_{\pm \mbox{ }\pm }\left( x\right) ,T_{\pm \mbox{ }\pm
}\left( y\right) \right\} &=&-8T_{\pm \mbox{ }\pm }\delta ^{\prime
}\left( x-y\right) -4T_{\pm \mbox{ }\pm }^{\prime ^{\prime
}}\delta \left(
x-y\right) ,  \label{ep5} \\
\left\{ T_{+\mbox{ }+}\left( x\right) ,T_{-\mbox{ }-}\left(
y\right) \right\} &=&0.  \label{ep6}
\end{eqnarray}

\section{Lax pair and the extended Yang-Baxter relations}
The field equations (\ref{s1})-(\ref{s2}) of the SPCM Lagrangian
(\ref{compt lagrangian}) are also obtained as the compatibility
condition of the following set of linear equations (Lax
pair)\cite{1}
\begin{eqnarray}
\partial _{+}V\left( x^{+},x^{-};\lambda \right) &=&A_{+}^{\left( \lambda
\right) }V\left( x^{+},x^{-};\lambda \right) ,  \nonumber \\
\partial _{-}V\left( x^{+},x^{-};\lambda \right) &=&A_{-}^{\left( \lambda
\right) }V\left( x^{+},x^{-};\lambda \right) ,  \label{linearsys}
\end{eqnarray}%
where $A_{_{\pm }}^{\left( \lambda \right) }=A_{_{\pm }}\left(
x^{+},x^{-};\lambda \right)$ is defined as,
\begin{equation}
A_{\pm }^{^{(\lambda )}}=\left\{ \mp \left( \frac{\lambda }{1\mp \lambda }%
\right) j_{\pm }+i\left( \frac{\lambda }{1\mp \lambda }\right)
^{2}h_{\pm }\right\} .  \label{s6}
\end{equation}%
The compatibility condition of the linear system (\ref{linearsys})
is the zero-curvature condition for the $\lambda -$dependent
connection components $A_{_{\pm }}^{\left( \lambda \right) } $
\begin{eqnarray}
&&\left[ \partial _{+}-A_{+}^{\left( \lambda \right) },\partial
_{-}-A_{-}^{\left( \lambda \right) }\right]  \nonumber \\
&\equiv &\partial _{-}A_{+}^{\left( \lambda \right) }-\partial
_{+}A_{-}^{\left( \lambda \right) }+\left[ A_{+}^{\left( \lambda
\right) },A_{-}^{\left( \lambda \right) }\right] =0.
\label{linear zc}
\end{eqnarray}%
Inserting from (\ref{s6}) in equation (\ref{linear zc}) gives%
\begin{eqnarray}
0 &=&-(\frac{\lambda }{1-\lambda })\partial _{-}j_{+}-(\frac{\lambda }{%
1+\lambda })\partial _{+}j_{-}+\frac{1}{2}\left( \frac{\lambda }{1-\lambda }-%
\frac{\lambda }{1+\lambda }\right) \times  \nonumber \\
&&\left( \left[ j_{+},j_{-}\right] -\frac{i}{2}\left[ j_{-},h_{+}\right] +%
\frac{i}{2}\left[ j_{+},h_{-}\right] -\frac{1}{2}\left[
h_{+},h_{-}\right]
\right)  \nonumber \\
&&+\left( \frac{\lambda }{1-\lambda }\right) ^{2}\left( i\partial _{-}h_{+}+%
\frac{1}{2}\left[ ih_{+},j_{-}+\frac{i}{2}h_{-}\right] \right)  \nonumber \\
&&-\left( \frac{\lambda }{1+\lambda }\right) ^{2}\left( i\partial _{+}h_{-}+%
\frac{1}{2}\left[ ih_{-},j_{+}+\frac{i}{2}h_{+}\right] \right) .
\label{comptzc}
\end{eqnarray}%
Since equation (\ref{comptzc}) holds for all values of $\lambda $ away from $%
\pm 1$, the coefficients of $\left( \frac{\lambda }{1-\lambda
}\right)
,\left( \frac{\lambda }{1+\lambda }\right) ,\left( \frac{\lambda }{%
1-\lambda }\right) ^{2}$ and $\left( \frac{\lambda }{1+\lambda
}\right)
^{2} $ must be separately zero. This gives equations (\ref{s3}) and (\ref{s5}%
) that are equivalent to the equations (\ref{s1})-(\ref{s2}). The
general
solution of the Lax pair (\ref{linearsys}) is%
\begin{equation}
V\left( x^{+},x^{-};\lambda \right) =e^{P\left(
x^{+},x^{-};\lambda \right) }V_{0}\left( \lambda \right) ,
\end{equation}%
where%
\begin{eqnarray*}
P\left( x^{+},x^{-};\lambda \right) &=&\frac{-\lambda }{1-\lambda }%
\int\limits_{x_{0}^{+}}^{x^{+}}j_{+}dy^{+}+i\left( \frac{\lambda
}{1-\lambda
}\right) ^{2}\int\limits_{x_{0}^{+}}^{x^{+}}h_{+}dy^{+} \\
&&+\frac{\lambda }{1+\lambda }\int\limits_{x_{0}^{-}}^{x^{-}}j_{-}dy^{-}+i%
\left( \frac{\lambda }{1+\lambda }\right)
^{2}\int\limits_{x_{0}^{-}}^{x^{-}}h_{-}dy^{-}.
\end{eqnarray*}%
In the above expression $V_{0}$ is the initial condition and is a
free element of the Lie group $\mathcal{G}.$ In terms of
space-time coordinates,
the associated linear system can be expressed as%
\begin{eqnarray}
\partial _{0}V\left( t,x;\lambda \right) &=&A_{0}^{\left( \lambda \right)
}V\left( t,x;\lambda \right) ,  \nonumber \\
\partial _{1}V\left( t,x;\lambda \right) &=&A_{1}^{\left( \lambda \right)
}V\left( t,x;\lambda \right) ,  \label{s-t linear system}
\end{eqnarray}%
with
\begin{eqnarray}
A_{0}^{\left( \lambda \right) } &=&\frac{-\lambda }{1-\lambda
^{2}}  \nonumber
\\
&&\left\{ j_{1}+\lambda j_{0}-\frac{i}{2}\lambda \left( \frac{1+\lambda }{%
1-\lambda }\right) h_{+}-\frac{i}{2}\lambda \left( \frac{1-\lambda }{%
1+\lambda }\right) h_{-}\right\} , \\
A_{1}^{\left( \lambda \right) } &=&\frac{\lambda }{1-\lambda ^{2}}  \nonumber \\
&&\left\{ j_{0}+\lambda j_{1}-\frac{i}{2}\lambda \left( \frac{1+\lambda }{%
1-\lambda }\right) h_{+}+\frac{i}{2}\lambda \left( \frac{1-\lambda }{%
1+\lambda }\right) h_{-}\right\} .  \label{A1}
\end{eqnarray}%
Using equations (\ref{CR1})-(\ref{CR7}) \ to find the Poisson bracket of $%
A_{1}$'s (the spatial part of the Lax pair) from (\ref{A1}) we get
\begin{eqnarray}
\left\{ A_{1}\left( x,\lambda\right) \stackrel{\otimes}{\mbox,}
A_{1}\left( y,\lambda\right) \right\} &=&\{\frac{-\lambda \mu
}{\left( 1-\lambda ^{2}\right) \left( \lambda -\mu \right) }\left[
c ,A_{1}\left( x,\lambda \right) \otimes
\mathbf{1}\right]  \nonumber \\
&&+\frac{-\lambda \mu }{\left( 1-\mu ^{2}\right) \left( \lambda
-\mu \right) }\left[ c ,\mathbf{1}\otimes A_{1}\left( x,\mu
\right) \right] \}\delta
\left( x-y\right)  \nonumber \\
&&+\frac{\lambda \mu \left( \lambda +\mu \right) }{\left(
1-\lambda ^{2}\right) \left( 1-\mu ^{2}\right) }c \delta ^{\prime
}\left( x-y\right) .  \label{PB algebra}
\end{eqnarray}%
The terms containing the brackets of $h_{\pm }$ cancel and we are
left with the terms which can be written in terms of the Lax
matrix $A_{1}\left( x,\lambda \right) $. In terms of $r$ and $s$
matrices we can rewrite the Poisson bracket as
\begin{eqnarray}
\left\{ A_{1}\left( x\right) \stackrel{\otimes}{\mbox,}
A_{1}\left( y\right) \right\} &=&\{\left[ (r-s)_{\lambda ,\mu
},A_{1}\left( x,\lambda \right)
\otimes \mathbf{1}\right]  \nonumber \\
&&+\left[ \left( r+s\right) _{\lambda ,\mu },\mathbf{1}\otimes
A_{1}\left(
x,\mu \right) \right] \}\delta \left( x-y\right)  \nonumber \\
&&-2s\left( \lambda ,\mu \right) \delta ^{\prime }\left(
x-y\right) , \label{pb}
\end{eqnarray}%
This result is very important, which on comparison with the
bosonic PCM \cite{16} shows that the Poisson bracket is of the
same form as that of the bosonic principal chiral model with no
independent contribution coming from the terms containing the
fields $h_{\pm }\left( x\right) $ and it reduces to the Poisson
bracket of bosonic model found in \cite{16}, when fermions are set
equal to zero. Since the Poisson bracket is same for both the
bosonic PCM and SPCM, therefore the algebra of monodromy matrices
obtained for the bosonic PCM \cite{5} can be extended to the case
of SPCM (see section 4). The matrices $r$ and $s$ are given as
\begin{eqnarray}
r\left( \lambda ,\mu \right) &=&\frac{-\lambda \mu }{2\left(
\lambda -\mu \right) }\left\{ \frac{1}{1-\mu
^{2}}+\frac{1}{1-\lambda ^{2}}\right\} c
\label{r1} \\
s\left( \lambda ,\mu \right) &=&\frac{-\lambda \mu \left( \lambda
+\mu \right) }{2\left( 1-\mu ^{2}\right) \left( 1-\lambda
^{2}\right) }c \label{r2}
\end{eqnarray}%
Equations $\left( \ref{r1}\right) $ and $\left( \ref{r2}\right) $
show that the $r$ and $s$ matrices obtained for the SPCM are same
as for the bosonic PCM. The antisymmetry of the canonical brackets
(\ref{pb}) holds through the relations
\begin{equation}
Pr\left( \lambda ,\mu \right) P=-r\left( \mu ,\lambda \right) ,\ \
\ Ps\left( \lambda ,\mu \right) P=s\left( \mu ,\lambda \right) ,
\label{yb4}
\end{equation}%
where $%
P_{ac,bd}=\delta _{ad}\delta _{cb}$ satisfies for any matrices $A,B;$ $%
P\left( A\otimes B\right) P=B\otimes A.$ It must be emphasized
that the algebra (\ref{pb}) is a non-trivial generalization of the
canonical
structure of those ultralocal type models which are obtained in the limit $%
s=0$ and $\partial _{x}r\left( x,\lambda ,\mu \right) =0$. Our algebra (\ref%
{pb}) is a linear algebra written in terms of two matrix structure
constants
($r$ and $s)$ with central extension $\left( \delta ^{\prime }\mbox{ term}%
\right) $ governed by the $s$-matrix. It is important to point out
here that in general the non-ultralocal integrable models exhibit
a space-time dependence for $r$ and $s$ matrices
\cite{13}-\cite{16} and there could be higher derivatives of the
delta function in the Poisson current algebra. In our case, the
supersymmetric model is of non-ultralocal type containing first
derivative of the delta function in its Poisson current algebra.
However, the $r$ and $s$ matrices do not contain space-time
dependence and therefore are non-dynamical. Such kind of
non-dynamical $r-s$ matrices also appear in the case of an
$SU\left( 2\right) $ WZW\ model as a non-ultralocal model
\cite{5}. The algebra can be expressed in a more transparent way
by
introducing Lax operator $\mathcal{D}\left( x,\lambda \right) ,$ defined as%
\begin{equation}
\mathcal{D}_{1}\left( x,\lambda \right) =\partial _{1}+A_{1}\left(
x,\lambda \right) ,
\end{equation}%
so that the Poisson bracket algebra can be equivalently expressed
in terms
of the differential operator $\mathcal{D}\left( x,\lambda \right) $ as%
\begin{eqnarray}
\left\{ \mathcal{D}_{1}\left( x,\lambda \right) \stackrel{\otimes}{\mbox,} %
\mathcal{D}_{1}\left( y,\mu \right) \right\} &=&-\left[ r\left(
x,\lambda ,\mu \right) \delta \left( x-y\right)
,\mathcal{D}_{1}\left( x,\lambda \right) \otimes
\mathbf{1}+\mathbf{1}\otimes \mathcal{D}_{1}\left( y,\mu
\right) \right]  \nonumber \\
&&+[s\left( x,\lambda ,\mu \right) \delta \left( x-y\right) ,\mathcal{D}%
_{1}\left( x,\lambda \right) \otimes \mathbf{1}  \nonumber \\
&&-\mathbf{1}\otimes \mathcal{D}_{1}\left( y,\mu \right) ].
\label{Loperator}
\end{eqnarray}%
Requiring now, the Jacobi identity of the canonical bracket
(\ref{pb}) to be satisfied, we get the following, extended
Yang-Baxter equation for the numerical $r$- and $s$-matrices (see
\cite{9})

\begin{eqnarray}
&&\left[ \left( r+s\right) _{13}\left( \lambda ,\eta \right)
,\left(
r-s\right) _{12}\left( \lambda ,\mu \right) \right]  \nonumber \\
&&+\left[ \left( r+s\right) _{23}\left( \mu ,\eta \right) ,\left(
r+s\right)
_{12}\left( \lambda ,\mu \right) \right]  \nonumber \\
&&+\left[ \left( r+s\right) _{23}\left( \mu ,\eta \right) ,\left(
r+s\right)
_{13}\left( \lambda ,\eta \right) \right]  \nonumber \\
&=&0.  \label{eyb51}
\end{eqnarray}%
Here the indices $1,2,3$ label the three spaces involved in
computing the algebra of three $A_{1}$-matrices and we have for
example $\left( r+s\right) _{12}\left( \lambda ,\mu \right)
=\left( r+s\right) \left( \lambda ,\mu \right) \otimes
\mathbf{1}_{3}.$ Again, (\ref{eyb51}) is a generalization of the
usual classical Yang-Baxter equation for $r$-matrices in
ultralocal type models which is obtained (in \ref{eyb51} as in
\ref{pb}) for $s=0$. However, we have to note that in going from
ultralocal type models to non -ultralocal ones, it is not
sufficient to simply add a central extension ($\delta ^{\prime }$
term) to the ultralocal algebra of $A_{1}$-matrices. In fact, it
is necessary, as can be seen from (\ref{pb}), to modify also the
$\delta \left( x-y\right) $ part with $s$ terms related to the
extension ($\delta ^{\prime }$ term) in order to satisfy the
Jacobi identity (\ref{eyb51}). Moreover, in general (\ref{eyb51})
holds with an $r$-matrix which itself does not satisfy the usual
classical Yang-Baxter equation for $r$-matrices
of ultralocal models, hence, showing the crucial role played by the new $s$%
-matrix. It is then possible, using (\ref{pb}) and (\ref{eyb51})
to derive the canonical algebra of two monodromy matrices in a
completely consistent manner, i.e., in agreement with the Jacobi
identity since (\ref{eyb51}) is verified.

\section{Algebra of monodromy matrices}
The monodromy matrix $T\left( x,y,\lambda \right) $ is defined in
terms of Lax matrix $A_{1}\left( x,\lambda \right) $ as

\begin{equation}
T\left( x,y,\lambda \right) =P\exp \int\limits_{y}^{x}A_{1}\left(
x^{\prime },\lambda \right) dx^{\prime }.  \label{i2}
\end{equation}%
The infinite volume limit of $T\left( x,y,\lambda \right) $ i.e.,

\begin{equation}
T\left( \infty ,-\infty ,\lambda \right) \equiv T\left( \lambda
\right) =P\exp \int\limits_{-\infty }^{\infty }A_{1}\left(
x,\lambda \right) dx, \label{i8}
\end{equation}%
is a conserved quantity for any value of spectral parameter $\lambda \cite%
{15}.$
 By expanding $%
T\left( \lambda \right) $ in powers of $\lambda $, an infinite set
of non-local conserved quantities is obtained with the first two
quantities given by \cite{1}-\cite{Curtright}$.$
\begin{eqnarray}
Q^{\left( 1\right) a} &=&-\int\limits_{-\infty }^{\infty }dy\mbox{ }%
j_{0}^{a}\left( t,y\right) ,  \label{q1} \\
Q^{\left( 0\right) a} &=&\int\limits_{-\infty }^{\infty }dy\mbox{ }%
[-j_{1}^{a}\left( t,y\right) +\frac{i}{2}\left( h_{+}^{a}\left(
t,y\right)
-h_{-}^{a}\left( t,y\right) \right)  \nonumber \\
&&+\frac{1}{2}f^{abc}j_{0}^{b}\left( t,y\right) \int\limits_{-\infty }^{y}dz%
\mbox{ }j_{0}^{c}\left( t,z\right) ].  \label{q11}
\end{eqnarray}%
The non-local conserved quantities generate a Yangian deformation symmetry $%
\cite{18}.$

The monodromy matrix usually contains the main information about
the canonical structure of the non-ultralocal sigma models. In
particular, its infinite volume limit (through proper
regularization) provides us when expanded in a power series in
$\lambda $, with an infinite set of conserved quantities, an
infinite subset of them being in involution i.e., Poisson commute,
as a signature of complete integrability of the model. Since the
Poisson bracket algebra of the $A_{1}$-matrices of the SPCM is
similar to that of the bosonic model, therefore the Poisson
bracket for the monodromy matrices of the SPCM can be determined
using the equal point limits through a regularization procedure
developed for the bosonic models in \cite{4}-\cite{19}$.$ The
Poisson bracket of the monodromy matrices of the SPCM turns out to
be of the same form as the Poisson bracket of the bosonic models
and is given by

\begin{equation}
\left\{ T\left( x,y,\lambda \right) \stackrel{\otimes}{\mbox,}
T\left( x,y,\mu \right) \right\} =\left[ r\left( \lambda ,\mu
\right) ,T\left( x,y,\lambda \right) T\left( x,y,\mu \right)
\right] . \label{t66}
\end{equation}

\smallskip\ In
the infinite volume
limit equation (\ref{t66}) reads%
\begin{equation}
\left\{ T\left( \lambda \right) \stackrel{\otimes}{\mbox,} T\left(
\mu \right) \right\} =\left[ r\left( \lambda ,\mu \right) ,T\left(
\lambda \right) T\left( \mu \right) \right] ,  \label{t70}
\end{equation}%
and the conserved quantities, Tr$T\left( \lambda \right) $ are in
involution, being
\begin{equation}
\mbox{Tr}\left( A\otimes B\right) =\mbox{Tr}A.\mbox{Tr}B\ ,
\end{equation}%
so that
\begin{equation}
\left\{ \mbox{Tr}T\left( \lambda \right) ,\mbox{Tr}T\left( \mu
\right)
\right\} =\mbox{Tr}\left\{ T\left( \lambda \right) \otimes %
T\left( \mu \right) \right\} =0.
\end{equation}%
In summary, we have calculated the Poisson bracket algebra of the $A_{1}$%
-matrices of the Lax pair of the SPCM as a non-ultralocal
integrable model. From the $A_{1}$-matrices of the SPCM, we have
determined the Poisson bracket algebra of the monodromy matrices
and using the equal-point limit, we have shown the existence of
conserved quantities of the model that are in involution with each
other establishing the classical integrability of the SPCM as a
non-ultralocal model. It seems appropriate here to make few
comments about the existence of an infinite number of conserved
quantities of the SPCM. It has been shown in \cite{1}, \cite{2}
that there exist an infinite number of nonlocal and local
conserved quantities of the SPCM. The nonlocal conserved
quantities can be generated from the monodromy matrix as has been
discussed in this section and they generate Yangian symmetry. The
local conserved quantities of the SPCM have been investigated in
\cite{2} and it has been shown that there are two families of
conserved quantities in involution, each with finitely many
members whose spins are the exponents of the underlying Lie
algebra. Similarly in \cite{1}, an infinite number of local
conservation laws has been constructed through a pair of matrix
Riccati equations of the SPCM. The appearance of these conserved
quantities has important consequences regarding the integrability
of the SPCM and they are constructed through a Lax pair and the
zero-curvature condition of the model. No explicit form of the
conserved quantities has been obtained either those of the trace
of monodromy matrix or those obtained through a set of matrix
Riccati equations. Once the explicit form of the conserved
quantities is known, one would be able to establish a relation
among these quantities
generated through different approaches.

\section{Conclusions}
We have developed an $r-s$ matrix formalism of the supersymmetric
principal chiral model as a non-ultralocal integrable model. By
evaluating the fundamental Poisson bracket of the $A_{1}$-matrices
of the Lax pair of the SPCM, we have shown that this bracket has
the same form as the fundamental Poisson bracket of the bosonic
principal chiral model. The fundamental Poisson bracket is then
used to define the monodromy matrix of the model, that gives the
conserved quantities in involution. The algebraic structures
studied here can also be investigated for the supersymmetric
nonlinear sigma models on Riemannian symmetric spaces that is the
most general class of
supersymmetric nonlinear sigma models to be integrable (see e.g. \cite%
{saleem2}, \cite{evans}). The other direction where the work can
be further extended is the recent investigations regarding the
classical integrability in superstring theory on the
$AdS_{5}\times S^{5}$ (see e.g. \cite{Bena}-\cite{29}). In these
studies, the theory has been regarded as a nonlinear sigma model
with the field
taking values in the supercoset space $\frac{PSU\left( 2,2/4\right) }{%
SO\left( 4,1\right) \times SO\left( 5\right) }$, which has an even part the $%
AdS_{5}\times S^{5}$ geometry. The even part admits a Lax
formalism and is further linked with conserved quantities of the
Yang-Mills sector of the AdS/CFT correspondence (see e.g.
\cite{Bena}-\cite{29} ). The algebra of monodromy matrices for the
$AdS_{5}\times S^{5}$ superstrings has been investigated in
\cite{Bena}-\cite{29}. In the light of our result, one can expect
that Poisson bracket algebra can be developed for the superstring
theory on $AdS_{5}\times S^{5}$ as a non-ultralocal theory that
gives conserved quantities in involution and it fits in the $r-s$
matrix formalism of integrable models. Another important direction
that can be pursued for future research is to develop an $r-s$
matrix formalism for the sigma models with target space
supersymmetry. The most important aspect of such investigations
is, however, to promote the classical integrability of such model
to the quantum level.

 {\large {\bf
{Acknowledgements}}}

One of the authors Bushra Haider would like to acknowledge the
enabling role of the Higher Education Commission, Pakistan and
appreciates its financial support through `` Indeginous 5000
fellowship program" for PhD studies in Science \& Technology.

\end{document}